# Requirements-driven Dynamic Adaptation to Mitigate Runtime Uncertainties for Self-adaptive Systems

Zhuoqun Yang
Academy of Mathematics and Systems Science
Chinese Academy of Sciences
Beijing, China
zhuoqun.y@hotmail.com

Wei Zhang, Haiyan Zhao, Zhi Jin
Institute of Software
School of EECS, Peking University
Beijing, China
{zhangw, zhhy}@sei.pku.edu.cn, zhijin@pku.edu.cn

***Abstract***: Self-adaptive systems are capable of adjusting their behavior to cope with the changes in environment and itself. These changes may cause runtime uncertainty, which refers to the system state of failing to achieve appropriate reconfigurations. However, it is often infeasible to exhaustively anticipate all the changes. Thus, providing dynamic adaptation mechanisms for mitigating runtime uncertainty becomes a big challenge. This paper suggests solving this challenge at requirements phase by presenting REDAPT, short for REquirement-Driven adAPTation. We propose an adaptive goal model (AGM) by introducing adaptive elements, specify dynamic properties of AGM by providing logic based grammar, derive adaptation mechanisms with AGM specifications and achieve adaptation by monitoring variables, diagnosing requirements violations, determining reconfigurations and execution. Our approach is demonstrated with an example from the Intelligent Transportation System domain and evaluated through a series of simulation experiments.

## I. Introduction

Self-adaptive systems are capable of adjusting their behavior to cope with the changes in environment and itself. These changes may cause runtime uncertainty, which refers to the system state of failing to achieve appropriate reconfigurations under requirements violation [1]. We divide the sources of runtime uncertainty into context uncertainty and components uncertainty. Generally, context uncertainty means the changes in execution environment, e.g. changes of bandwidth for a service system or changes of temperature for an air conditioner, while components uncertainty represents changes in the system itself, e.g. sensors break for components-based systems. For keeping continuous satisfaction of requirements at runtime, self-adaptive systems need appropriate adaptation mechanisms. This paper suggests achieving adaptation mechanisms at requirements phase.

Requirements phase is considered as the first stage during the life cycle of a system. Definitely different from traditional requirements engineering (RE), for self-adaptive systems, it is need to capture not only functional requirements (FR) and non-functional requirements (NFR) but also adaptive requirements (AR), which refer to the requirements that can be hold through adaptation. Additionally, adaptation mechanisms also should be taken into consideration at this phase for making clear when to adapt and how to adapt [2]. Therefore, requirements engineering for self-adaptive systems should provide two kinds of support: methods for modeling requirements and adaptation mechanisms for mitigating runtime uncertainty. Recently, the RE community made great strides in introducing methods and techniques for providing the two supports. Wittle et al. [4] introduced RELAX, a formal requirements specification language to specify the uncertain requirements of self-adaptive systems. Cheng et al. extended RELAX with goal modeling to specify uncertainty in the goal model [5]. Requirements-aware approaches [11-14] considered requirements as runtime entities and monitoring as meta-requirements about whether the requirements hold during runtime. Baresi et al. [16] introduced adaptive goal into KAOS model and provided adaptation mechanism by adopting fuzzy membership function. Other works include monitoring and diagnose [8-10], self-repair [17] [18] and architecture-based adaptation [20-23].

However, on account of growing complexity of system structure, inherent volatility of environment and increasing diversity of seamless interaction between the system and environment, it becomes infeasible to predict and anticipate all the runtime changes at requirements phase. Moreover, in more and more situations, changes cannot be handled off-line, but require the system to adapt its behavior dynamically without human intervention. Therefore, providing requirements models at runtime for resolving unpredictable uncertainty and model-driven adaptation mechanisms become research urgencies and challenges [24] [25].

This paper tackles these challenges by seeking answers to the following questions: how to identify the boundary of uncertainty sources (Q1); how to provide runtime requirements model for modeling adaptive requirements and uncertainty sources in a generic process (Q2); how to represent dynamic properties of the runtime requirements model (Q3); how to derive adaptation mechanisms and achieve adaptation (Q4).

To achieve these ends, we endeavor to present REDAPT, short for REquirements-Driven adAPTation. For answering Q1, this paper introduces how to exploring problem world for identifying the boundary of uncertainty sources. To answer Q2, we propose an adaptive goal model (AGM) by introducing context uncertainty, components uncertainty, kinds of adaptive elements and MAPE loop [26] into traditional goal model. For answering Q3, we provide first-order linear-time temporal logic based formal grammar for specifying AGM elements so as to bridge the gap between requirements model and adaptation mechanisms. To answer Q4, we derive adaptation mechanism algorithms by integrating AGM with its specifications. The adaptation can be achieved by monitoring variables, diagnosing requirements violations, determining reconfigurations and execution. We demonstrate REDAPT by applying it to a Highway-Rail Control System (HRCS) from Intelligent Transportation System domain (ITS) and evaluate it through a series of simulation experiments. The results illustrate REDAPT's ability of modeling requirements as runtime entities and mitigating runtime uncertainty by parametric adaptation and structural adaptation.

The rest of the paper is structured as follows. Section II introduces the motivating example. Section III overviews our approach. Section IV presents the processes of modeling AGM and deriving its specifications. Section V provides the adaptation mechanism algorithm. We illustrate and evaluate the proposed adaptation mechanism in Section VI. The related work is discussed in Section VII. Finally, Section VIII concludes the paper and identifies avenues for future work.

## II. Motivating Example

Intelligent transport systems are advanced applications which, without embodying intelligence as such, aim to provide innovative services relating to different modes of transport and traffic management and enable various users to be better informed and make safer. Highway-rail crossing control is a fundamental concern [29]. Basic scenarios can be described as:

- Rails are built across a highway for trains dispatched from east and west. At the crossing, gates are built at both sides of the highway for block the vehicle flow temporarily from south and north when train is coming.
- *Train Dispatch End* is in charge of determining appropriate dispatch time interval according to vehicle flow.
- *Gate Control End* is in charge of closing/opening gates on both side of the highway, when the train is monitored approaching/leaving the crossing.

However, the environment and system itself are ever changed. Thus, the changing scenarios can be described as:

- When illuminance is above 20lx, the closing/opening time interval is set to 4s for pass efficiency; otherwise the closing time interval should be set within (1s, 4s] and opening time interval should be set within [4s, 7s) for safety efficiency.
- Vehicle flow on highway is changing. For preventing traffic jam and accident, we add some constraints that the dispatch time interval should ensure at least 50% of the vehicles can pass through the whole highway within 400s and the amount of vehicles on the highway is always under 350.
- Sometimes sensors may fail to monitor context variables or the monitored value may suffer from noises.

The scenarios above imply that the dispatch time interval is not fixed but varies according to the changed vehicle flow. Thus, the requirement of determining dispatch time interval should adapt to vehicle flow changes. We call this kind of requirement Adaptive Requirements (AR). Adaptive requirements of HRCS are described in Table I.

TABLE I. Adaptive Requirements of HRCS

| AR | Description |
|---|---|
| AR1 | When illuminance is above 20lx, gates open/close time interval is set to 4s for pass efficiency. |
| AR2 | For safety efficiency, when illuminance is under 20lx, gates' closing time should be set within (1s, 4s], while opening time should be set within [4s, 7s) to achieve safety efficiency. |
| AR3 | Dispatch time interval should ensure at least 50% of the vehicles can pass through the whole highway within 400s. |
| AR4 | Dispatch time interval should ensure the amount of vehicles on the highway is always under 350 at any time. |
| AR5 | Components can be replaced once failure occurs. |

## III. Approach Overview

This section provides the framework of REDAPT and the processes of deriving adaptive goal model, AGM specifications and achieving adaptation.

### A. REDAPT Framework

Figure 1 depicts the REDAPT framework. It is composed of three basic layers: Problem Layer, Model Layer and Adaptation Layer. Model Layer models the scenarios from Problem Layer, while Adaptation Layer achieves the adaptation mechanisms derived from Model Layer. The three layers compose a feedback loop that controls the adaptation of self-adaptive systems at runtime.

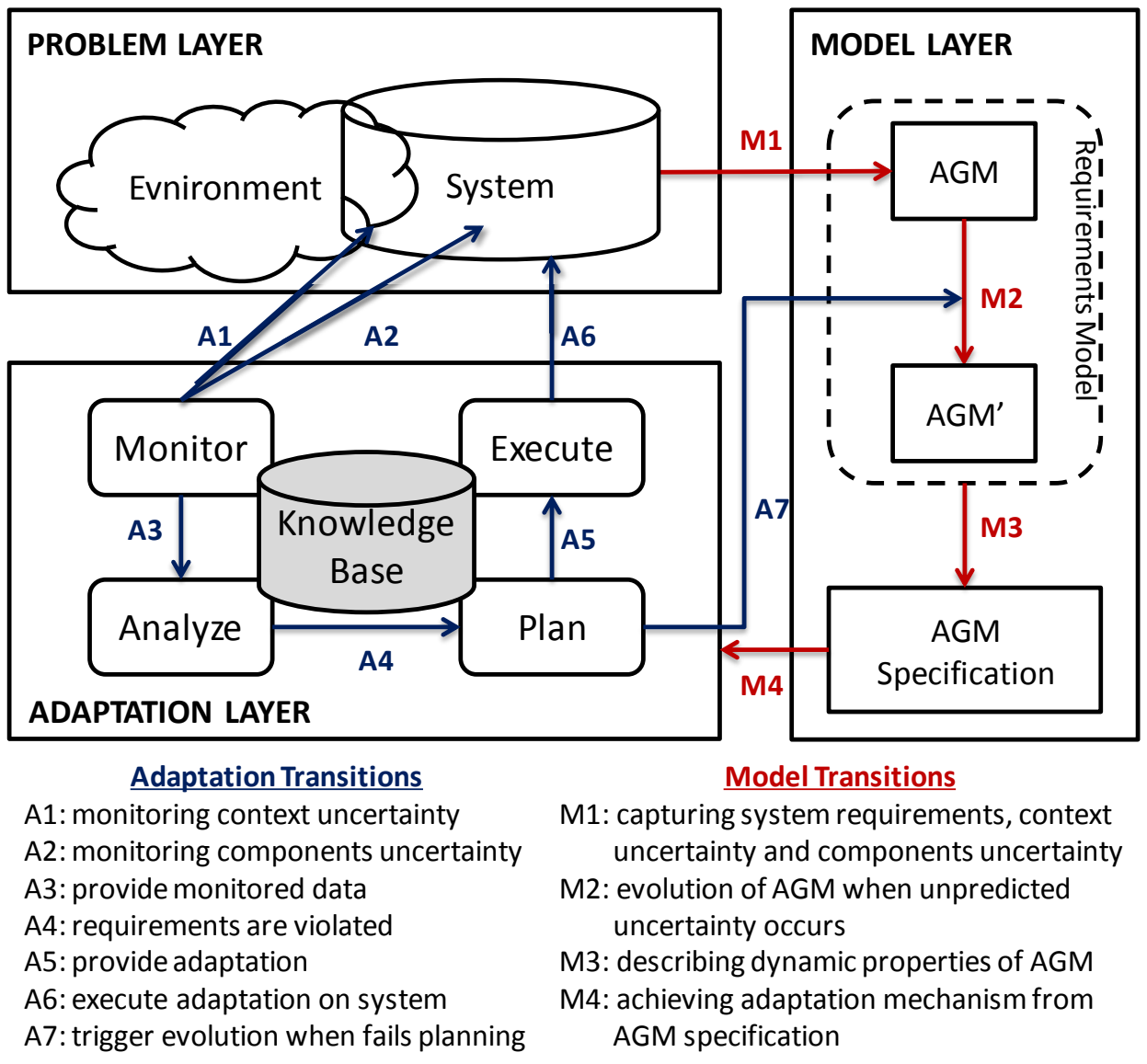

Fig. 1. Overview of the REDAPT framework

Problem layer aims at illustrating the relation between the system and the environment. By analyzing the shared phenomena and system, we can capture system's requirements, context uncertainty and components uncertainty (M1).

Model layer consists of Adaptive Goal Model (AGM) and AGM Specifications. AGM is the requirements model derived by extending traditional goal model with some new elements, e.g. context uncertainty, adaptive goal and adaptive task. We consider AGM as runtime entity, which can evolve at runtime (M2). For bridging the gap between AGM and adaptation mechanisms, dynamic properties of AGM are represented with AGM specifications (M3). Thus we can achieve adaptation mechanism algorithms from the specifications (M4).

Adaptation layer presents adaptation mechanisms based on AGM specification. We adopt the MAPE feedback loop from autonomic computing [26] for monitoring context uncertainty (A1) and components uncertainty (A2), diagnosing requirements violations (A4), deciding resolutions (A5) and executing the decisions (A6). When the system fails planning, unpredicted uncertainty should be modeled in AGM (A7).

### *B. From Model Layer to Adaptation Layer*

#### *1) Deriving Adaptive Goal Model and Specifications*

Figure 2 presents a generic process of deriving AGM and its specifications in Model Layer. The input is scenarios and the output is AGM specifications. Each sub-process connects with each other by artifacts coming from the former process. Directed lines depict the order of processes. We give a brief introduction to these sub-processes next.

P1: Adaptive requirements are elicited from the scenarios and we can derive the initial requirements model in P1. The adaptive requirements are described in adaptive tasks.

P2: Context uncertainty and components uncertainty are identified from the adaptation scenarios.

P3: Attach context uncertainty to the adaptive tasks in existing requirements model and then switch the adaptive tasks into adaptive goals.

P4: Refine adaptive goals with MAPE loop. Thus, each adaptive goal is operationalized through four tasks: Monitor, Analyze, Plan and Execute.

P5: Attach components uncertainty to the tasks that are accomplished by components and switch the tasks into adaptive goals again.

P6: Refine new adaptive goals with MAPE tasks again.

P7: Specify elements of AGM with proposed grammar.

AGM can be derived from P1 to P6. P7 is involved in the process because it has close relation with AGM. The details of specifications are elaborated in Section IV.

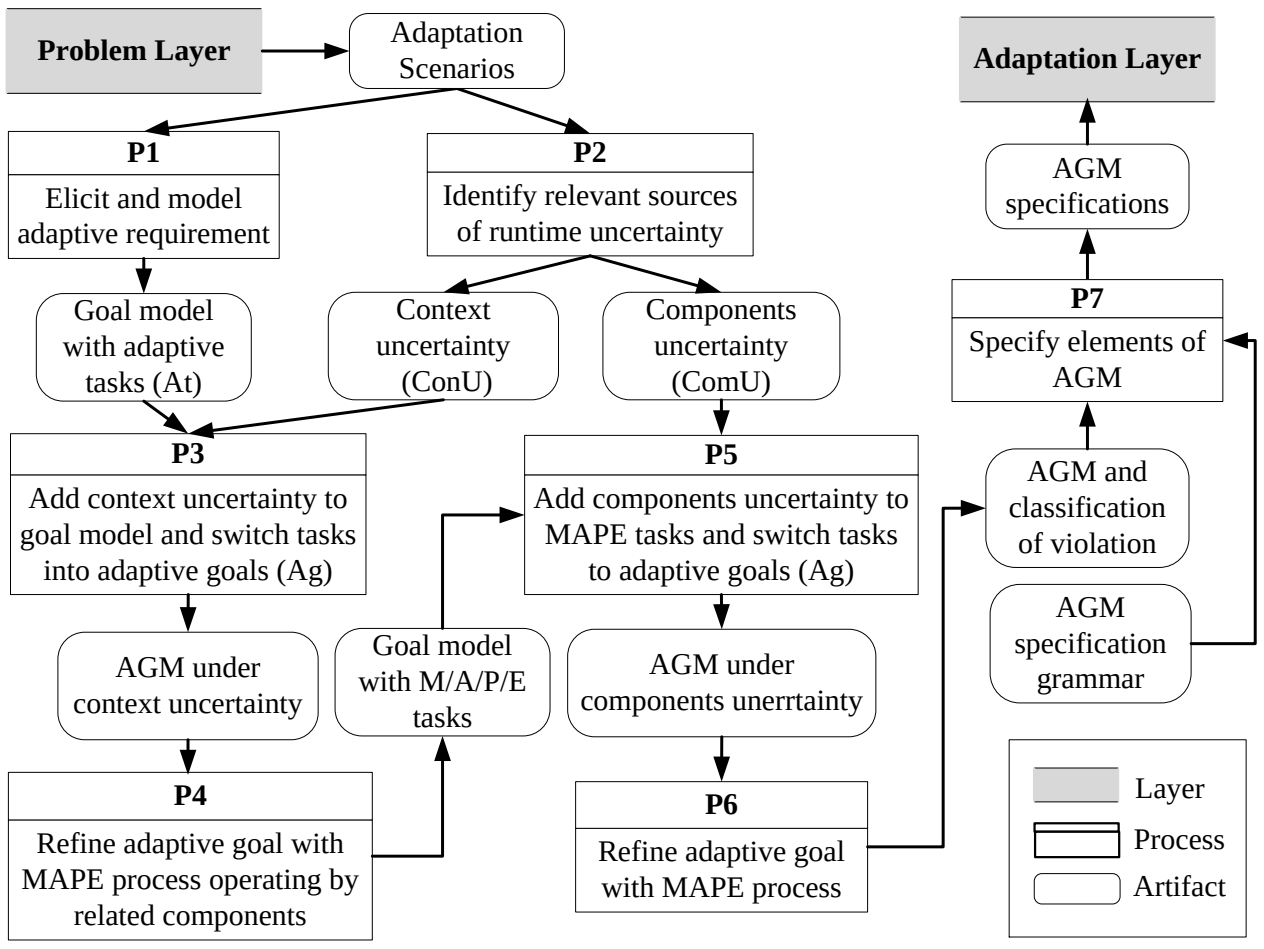

Fig. 2. A generic process of deriving adaptive goal model

### *C. Process of Achieving Requirements-driven Adaptation*

Figure 3 depicts a generic process of achieving adaptation in Adaptation Layer. The input is AGM specifications and the output is new configurations.

Monitoring: Monitor algorithms are derived from specifications of Monitor tasks. The monitored variables are specified in attributes of monitor tasks. By monitoring, we get runtime data of both context and components.

Analyzing: Specifications of Analyze tasks describe how to diagnose requirements violation at runtime. We suggest adopting quantitative verification into diagnosing process. For choosing appropriate algorithms, we should match the type of violation according to the classification derived from both requirements perspective and uncertainty perspective.

Planning: Planning algorithms should be chosen according to the type of violation. What should be noticed is that the planning process is intertwined with analyzing process, because the modification of parameters or structures is always followed by iterative verification.

Executing: Realize parametric or structural adaptation according to the results from planning process.

After adaptation, new configuration is achieved and runtime uncertainty is mitigated. The achieving details are provided in Section V.

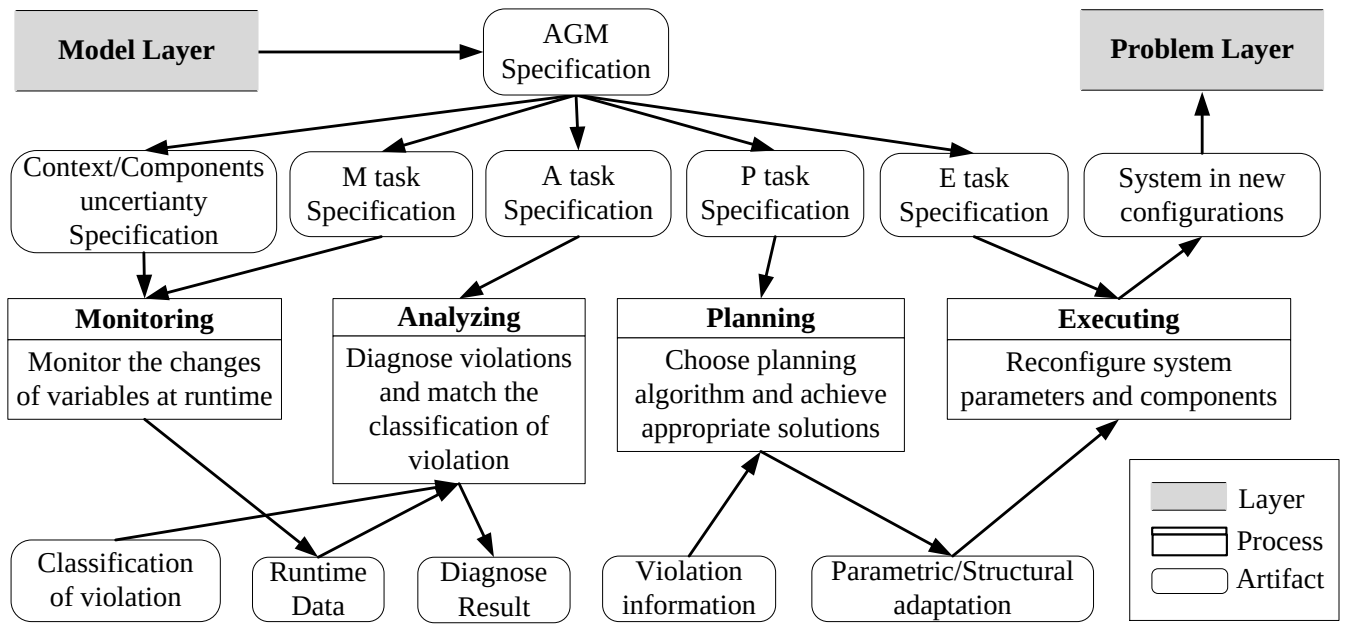

Fig. 3. A generic process of achieving requirements-driven adaptation

## IV. ADAPTIVE GOAL MODEL AND FORMAL SPECIFICATION

REDAPT provides the processes of deriving AGM and its specifications. First, we should make clear the boundary of uncertainty and then carry out the processes in Section IV-B.

### *A. Identifying Uncertainty*

#### *1) Capturing Uncertainty*

Uncertainty should be identified according to concrete scenarios. Problem Layer in Section III consists of two intertwined parts: system and environment. The uncertainty existing in environment that can be observed by system belongs to context uncertainty. While, the uncertainty existing in system itself belongs to components uncertainty. Both context uncertainty and components uncertainty are unpredicted at runtime.

In HRCS, context uncertainty consists of:

- *Illuminance*: Outdoor illuminance is a continuous variable changing over time. 20lx is a boundary of whether driver can drive safely without auxiliary illumination instruments or aids. Thus, when illuminance is under 20lx, the system should adjust it behavior for safety efficiency requirement.

- *Vehicle flow*: Vehicle flow can be viewed as discrete variable because the monitored results are always integer.

Components uncertainty consists of:

- *Sensor Failure*: Sensors fails to monitor the changes in context. There is no return value from sensors. Failed sensors should be replaced by available sensors.
- *Sensor Noise*: Sensors succeed in monitoring variables, but the monitored value is unstable. Sensor noise can also tackled by replacement of available sensors.

*2) Symbol Assumptions*

For the convenience of deriving AGM and specifications, Table II introduces some symbols to represent variables and constants in the adaptive scenarios. Utility is used to depict the satisfaction degree of requirements, varying from 0 to 1. 0 refers to completely dissatisfaction while 1 refers to completely satisfaction. The value between (0, 1) refers to partial satisfaction. For instance, $U_{close}$ represents the drivers' satisfaction degree of closing gates when train is coming. The small $t_{close}$ is, the small $U_{close}$ will be, because drivers want to pass the crossing as soon as possible. On the contrary, the small $t_{open}$ is, the large $U_{open}$ will be, because drivers don't want to wait at the crossing for too long. Utility is used to get better tradeoff between NFR and mitigate the runtime uncertainty of NFR. Among the symbols, $U_{close}$ is the function of $t_{close}$, while $U_{open}$ is the function of $t_{open}$. $U_{safety}$ is the function of $U_E$, $U_{close}$ and $U_{open}$, while $U_{pass}$ is the function of $U_{close}$ and $U_{open}$. We will give the concrete function in Section VI.

TABLE II. SYMBOL ASSUMPTIONS OF VARIABLES

| **Symbol** | **Meaning** |
|---|---|
| $e_i$ | Value of illuminance gauged by $I_sensor_i$ |
| $f_i$ | Value of vehicle flow gauged by $I_camera_i$ |
| $E$ | Mean value of illuminance |
| $F$ | Mean value of vehicle flow at each highway entrance |
| $t$ | Time |
| $t_{dispatch}$ | Dispatching time interval |
| $t_{close}$ | Time interval of gate close after detecting train's coming |
| $t_{open}$ | Time interval of gate opening after detecting train's leaving |
| $U_{close}$ | Utility of closing time for passing vehicles, proportional to $t_{close}$ |
| $U_{open}$ | Utility of opening time for passing vehicles, inversely proportional to $t_{open}$ |
| $U_E$ | Utility of illuminance |
| $U_{safty}$ | Utility of safety efficiency at crossing |
| $U_{pass}$ | Utility of pass efficiency at crossing |
| $p$ | Percentage of vehicles whose driving time is under 300 |
| $n$ | Number of vehicles on highway at each moment |

## *B. Deriving Adaptive Goal Model*

Goal-oriented methods are wildly accepted in RE [27]. Goal models are used to model and analyze stakeholder objectives. Systems' FR are represented as hard goals, while NFR are represented as softgoals. Figure 1 presents an example of goal model. Goal can be refined through AND/OR decompositions into sub-goals and sub-tasks. For AND decomposition, father goal can be achieved by achieving all the sub-goals or sub-tasks, while for OR decomposition it can be achieved by achieving only one of the alternative goals or tasks. For example, g2 can be achieved by achieving both t1 and t2, while g4 can be achieved by achieving either t4 or t5. Tasks are leaves in goal model and can be operated by certain agents. Tasks can contribute to softgoals through Help-contribution or Hurt-contribution. For instance, t4 has positive effects on sg1, while t3 has negative effects on sg1.

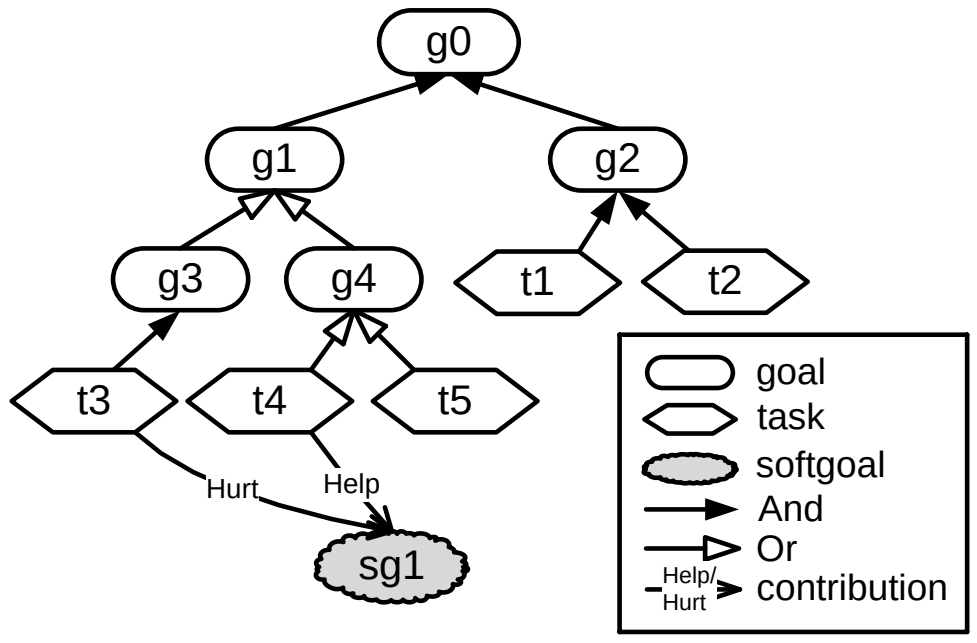

Fig. 4. An example of goal model

Figure 5 presents the goal model with adaptive tasks (*At*) derived through P1 in Section IV-B. Adaptive tasks model and represent the adaptive requirements. In general, *At1* models AR1 and AR2, while *At2* models AR3 and AR4. P2 is the process of identifying context uncertainty (*ConU*) and components uncertainty (*ComU*), which is accomplished in Section V-A. Next, we conduct the processes from P3 to P6 for both *At1* and *At2*. First we take *At1* for illustration.

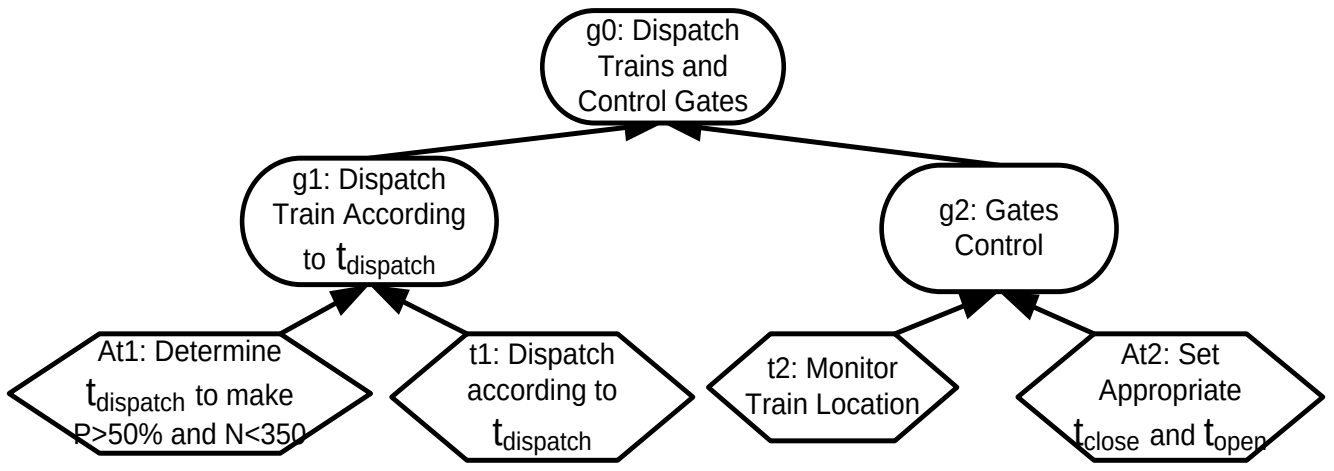

Fig. 5. Goal model with adaptive tasks

Figure 6 provides the processes of refining *At1*. Vehicle Flow, represented as *ConU1*, belongs to context uncertainty and it affects the achievement of *At1*. Thus, for P3, link *ConU1* to *At1* with *Affect* relation, represented as two-arrow directed line, attached with the label of FR violation. For refining *At1*, switch adaptive task into adaptive goal (*Ag*), i.e. *Ag1*. For P4, *Ag1* is decomposed into four tasks (*M1*, *A1*, *P1* and *E1*), which are adopt from MAPE loop. Thus, the original adaptation mechanisms are presented as M/A/P/E tasks. Similar processes can be carried out for modeling *At2* (Figure 7).

After modeling all the context uncertainty, we take into consideration of the components uncertainty (P5). *M1* can be achieved by Infrared Sensors, which may be affected by Sensor Failure (*ComU1*) and Sensor Noise (*ComU2*). *ComU1* may result in FR violation, while *ComU2* may cause NFR violation. Similar to *At1*, we switch *M1* into *Ag3* and refine it. Thus we derive the original adaptation mechanisms for achieving *Ag3*.

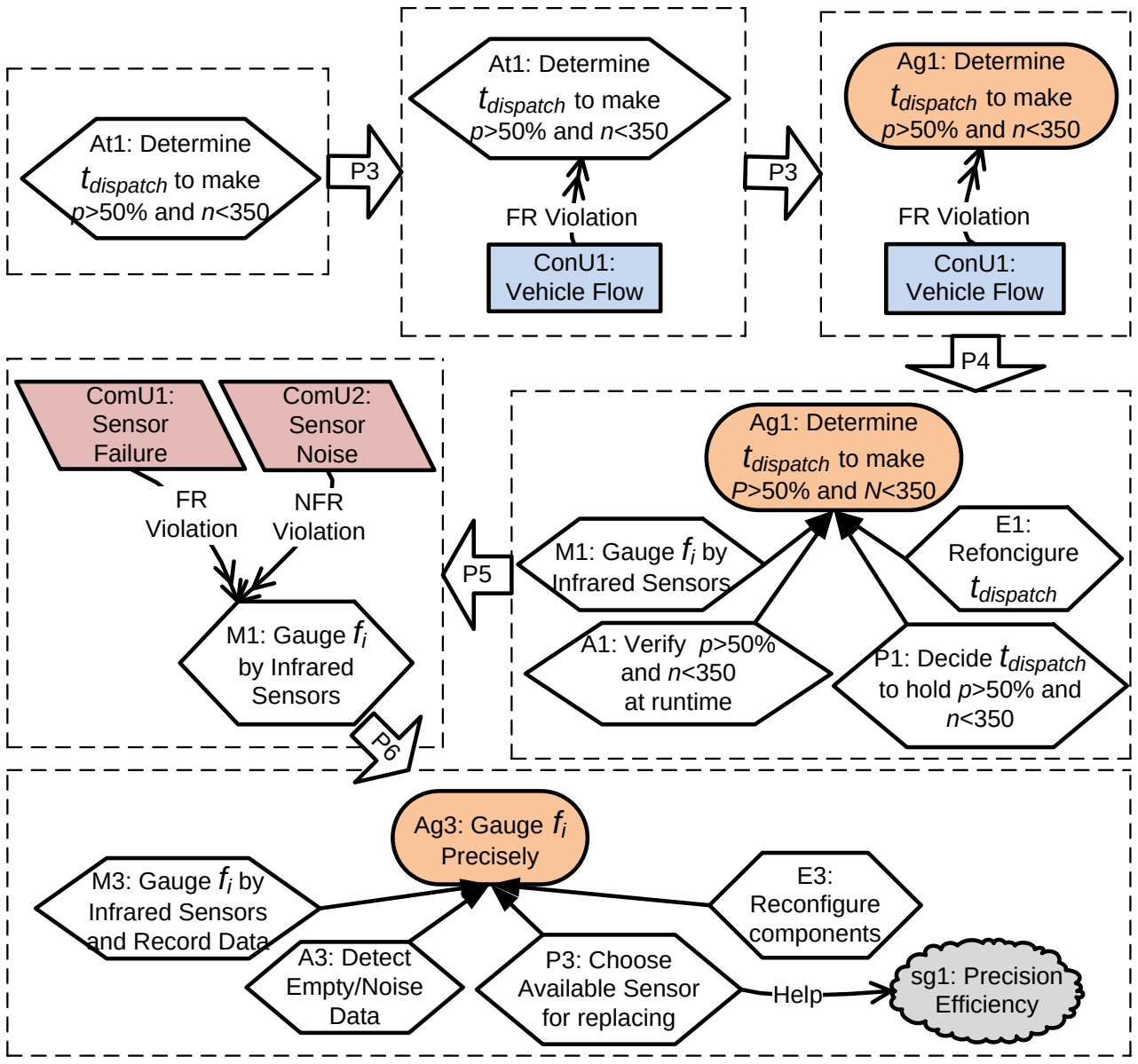

Fig. 6. Refinement of *At1*

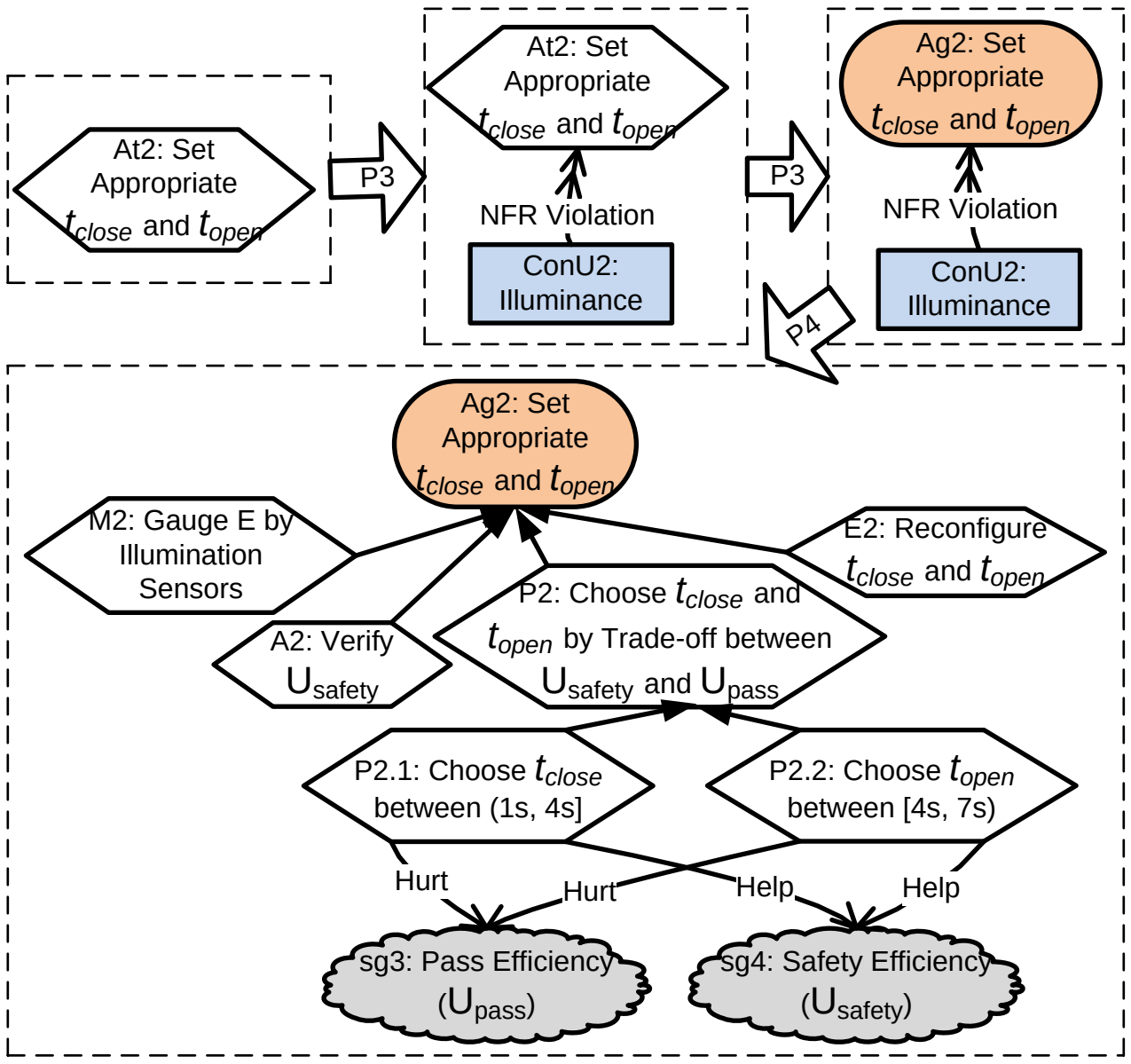

Fig. 7. Refinement of *At2*

After deriving the AGM, we should classify the violations for the convenience of choosing appropriate adaptation mechanisms during runtime. According to the *Affect* relation in Section V-B, we suggest classifying violations into four kinds from both uncertainty perspective and requirements perspective. Table III presents the classification of requirements violations of HRCS. Based on the four kinds of violation, we can design respective adaptation mechanism algorithms in Section VI. Among the four kinds of violations, components uncertainty caused violations can be solved through structural adaptation. While context uncertainty caused violations can be mitigating by parametric adaptation.

TABLE III. Classification of Violations

| **Requirements Type** | **Uncertainty Type** | |
|---|---|---|
| | ***Context Uncertainty*** | ***Components Uncertainty*** |
| Functional Requirements | Context uncertainty caused FR violation (violation of *Ag1*) | Components uncertainty caused FR violation (violation of *Ag3*) |
| Non-functional Requirements | Context uncertainty caused NFR violation (violation of *Ag2*) | Components uncertainty caused NFR violation (violation of *Ag3*) |

### *C. Adaptive Goal Model Specificationsl*

For representing the dynamic properties of AGM and deriving adaptation mechanisms later, we provide AGM specification's grammar in Figure 8, inspired by Formal Tropos [27] and KAOS specifications [28].

**// ELEMENTS**
*entity* := *goal* | *softgoal* | *task* | *uncertainty*
*goal* := *goal-type mode name* [*attributes*] [*initialization*]
  [*invariant*] [variant] [*fulfillment*]
*softgoal* := **Softgoal** *name* [*attribute*] [*tradeoff-softgoal*]
  [*invariant*] [variant] [*fulfillment*]
*task* := *task-type name* **From** *goal-type name input output*
  [*attributes*] [*initialization*] [*fulfillment*]
*uncertainty* := *uncertainty-type name* [*attribute*] [*violation*]
*goal-type* := **Ordinary Goal** | **Adaptive Goal**
*mode* := **achieve** | **maintain**
*tradeoff-softgoal* := **Tradeoff With Softgoal** *name*
*task-type* := **Ordinary Task**| **Monitor** | **Analyze** | **Plan** | **Execute**
*input* := **Input** *name*
*output* := **Output** *name*
*uncertainty-type* := **Context Uncertainty** | **Components Uncertainty**
*violation* := **Affected Adaptive Goal** *name formula violation-type*
*violation-type* := **FR Violation**| **NFR Violation**
**// ATTRIBUTES**
*attribute* := **Attribute** *attribute*$^+$
*attribute* := *attribute-type* : *name*
*attribute-type* := Numeric | Boolean | Class
**// INVARIANT, VARIANT, INITIALIZATION, FULFILLMENT**
*initialization* := **Initialization** *initial-property*$^+$
*initial-property* := *property-type conditional-type formula*
*invariant* := **Invariant** *invar-property*$^+$
*invar-property* := **Constraint** *formula*
*variant* := **Variant** *var-property*$^+$
*var-property* := **Possibility** *formula*
*fulfillment* := **Fulfillment** *fulfill-property*$^+$
*fulfill-property* := *property-type conditional-type formula*
*condition-type* := **PreCondition** | **TriggerCondition** | **PostCondition**

Fig. 8. AGM specification grammar

AGM specifications consist of specifications of entities, i.e. goal, softgoal, task and uncertainty. *Attribute* presents some properties related to the *entity*. *Attribute-type* can be Numeric, Boolean or Class. Numeric attributes depicts the variables and constants that are needed to achieving the entity. Boolean attributes always function as the output of verification activity. Class attributes refer to other entities involved during achieving the specified entity. For example, *attribute* of *M3* (Gauge F by Infrared Sensors) contains numeric attribute *F* (Vehicle Flow) and Class attribute I_sensor (Infrared Sensor). *Attribute* of *A1* contains a Boolean to record the result of verifying R3 and R4. Besides, father goals' attributes contain sub-goals' attributes and sub-tasks' attributes.

For goal, softgoal and task, *initialization* and *fulfillment* refer to the activating process and the terminating process of an entity respectively. *Condition-type* consists of PreCondition, TriggerCondition and PostCondition. PreCondition means the condition before initialization or fulfillment; TriggerCondition means the trigger condition; PostCondition means the result after initialization or fulfillment. It is known that father entities can be satisfied by achieving child entities. That is to say, child entities' initialization conditions are always triggered by the activation of their father entities, while father entities' fulfillment conditions are always triggered by fulfillment of their child entities. For instance, *Ag1*'s initialization is triggered by activation of *g1*, while its fulfillment is triggered by fulfillment of *M1*, *A1*, *P1* and *E1*.

*Invariant* refers to the constraints that the entities should hold all the time, while variant refers to the possibilities that may occur. For example, R3 and R4 are constraints to *Ag1*, while the priority of *sg2* and *sg3* is changeable. *Violation* is unique in *uncertainty*, describing the affected adaptive goals and violation types.

The formulas in the grammar are specified by first-order linear-time temporal logic. The syntax is given by:

$$
\begin{aligned}
t &::= x \mid c \mid \mathrm{f}(t_1,\ldots,t_n) \\
\phi &::= t \mid \mathrm{P}(t_1,\ldots,t_n) \mid \\
&\quad \neg\phi \mid \phi \wedge \phi \mid \phi \vee \phi \mid \phi \rightarrow \phi \mid \\
&\quad \mathrm{X}\,\phi \mid \mathrm{F}\,\phi \mid \mathrm{G}\,\phi \mid \phi\ \mathrm{U}\ \phi \\
&\quad \forall x \cdot \phi \mid \exists x \cdot \phi
\end{aligned}
$$

A term t is a variable x, a constant c, or a function f of a number of terms. A formula $\Phi$ is either a term, a predicate of a number of terms, a Boolean operation, a timed operation or a quantifiers operation. X$\Phi$ refers to $\Phi$ should hold in the next state reached by the system. F$\Phi$ refers to $\Phi$ should eventually hold in some future state. G$\Phi$ refers to $\Phi$ should hold in all state of the system. $\Phi_1$ U $\Phi_2$ refers to $\Phi_1$ should hold until $\Phi_2$ holds.

The grammar provides a template for representing dynamic properties of model entities. Several examples are provided.

**Context Uncertainty** Vehicle Flow
**Attribute**
**Numeric** $f_i$, $F$
**Class** I_sensor
**Affected Adaptive Goal** Determine $t_{dispatch}$ to make $p$>50% and $n$<350
$\exists F \cdot (p(t_{dispatch}, F) < 50\% \vee n(t_{dispatch}, F) > 350)$
**FR Violation**
**Components Uncertainty** Sensor Failure
**Attribute**
**Numeric** $f_i$
**Class** I_sensor
**Affected Adaptive Goal** Gauge $f_i$ Precisely
$\exists I_sensor_i \cdot I_sensor_i.value = ""$
**FR Violation**

Fig. 9. Specification of *ConU1* and *ComU1*

**Adaptive Goal Achieve** Determine $t_{dispatch}$ to make $p$>50% and $n$<350
**Attribute**
**Numeric** $t_{dispatch}$, $f_i$, $F$, $p$, $n$
**Class** I_sensor
**Initialization PreCondition**
$\neg Fulfill($ Dispatch Train According to $t_{dispatch})$
**Initialization TriggerCondition**
$Activate($ Dispatch Train According to $t_{dispatch})$
**Initialization PostCondition**
$t_{dispatch} = ""$
**Invariant**
$G(p \geq 50\%_0 \wedge n \leq 350)$
**Fulfillment PreCondition**
$\exists t_{dispatch} \cdot \mathrm{F}\,(p(t_{dispatch}, F) \geq 50\% \wedge n(t_{dispatch}, F) \leq 350)$
**Fulfillment TriggerCondition**
$\exists t^{new}_{dispatch}\ (p(t^{new}_{dispatch}, F) \geq 50\% \wedge n(t^{new}_{dispatch}, F) \leq 350)$
**Fulfillment PostCondition**
$t_{dispatch} = t^{new}_{dispatch}$

Fig. 10. Specification of *Ag1*

**Softgoal Maintain** Safety Efficiency
**Attribute**
**Numeric** $t_{close}$, $t_{open}$, E, $\mathrm{U_{safety}}$, $\mathrm{U_{pass}}$
**Tradeoff With Softgoal** Pass Efficiency
**Invariant Constraint**
$\neg\exists\{t_{close}, t_{open}, E\} \cdot \mathrm{G}(\mathrm{U_{safety}}(t_{close}, t_{open}, \mathrm{E}) \leq 0)$
**Variant Possibility**
$\forall t \cdot (\mathrm{Prior}(\textit{safety efficiency}, \textit{pass efficiency}, t) \vee$
$\mathrm{Prior}(\textit{pass efficiency}, \textit{safety efficiency}, t) \vee$
$\mathrm{equalPrior}(\textit{pass efficiency}, \textit{safety efficiency}, t))$
**Fulfillment PreCondition**
$\exists\{t_{close}, t_{open}\} \cdot \mathrm{U_{safety}}(t_{close}, t_{open}, \mathrm{E}) < \mathrm{U^{desired}_{safety}}$
**Fulfillment TriggerCondition**
$\exists\{t^{new}_{close}, t^{new}_{open}\} \cdot (U_{safety}(t^{new}_{close}, t^{new}_{open}, \mathrm{E}) \geq U^{desired}_{safety}) \wedge$
$MAX_Tradeoff(U_{safety}, U_{pass})$
**Fulfillment PostCondition**
$t_{close} = t^{new}_{close} \wedge t_{open} = t^{new}_{open}$

Fig. 11. Specification of sg2

**Monitor** Gauge $f_i$ by Infrared Sensors
**From Adaptive Goal** Determine $t_{dispatch}$ to make $p$>50% and $n$<350
**Attribute**
**Numeric** $f_i$
**Class** I_sensor
**Input** None
**Output** $f_1 \ldots f_{10}$
**Initialization PreCondition**
$\neg Fulfill(Determine\ t_{dispatch}\ to\ make\ p > 50\%\ and\ n < 350)$
**Initialization TriggerCondition**
$activate(Determine\ t_{dispatch}\ to\ make\ p > 50\%\ and\ n < 350)$
**Initialization PostCondition**
$\exists I_sensor_1, \ldots I_sensor_{10} \cdot I_sensor_i.select = \mathrm{TRUE}$
**Fulfillment PreCondition**
$\forall I_sensor_i, i \in \{1,\ldots,10\} \cdot I_sensor_i.value = ""$
**Fulfillment TriggerCondition**
$\forall I_sensor_i, i \in \{1,\ldots,10\} \cdot I_sensor_i.gauge = \mathrm{TURE}$
**Fulfillment PostCondition**
$I_sensor_i.value = guagedValue \wedge output(I_sensor_i.value)$

Fig. 12. Specification of *M1*

**Analyze** Verify *p*>50% and *n*<350 at runtime
**From Adaptive Goal** Determine $t_{dispatch}$ to make *p*>50% and *n*<350
**Attribute**
**Numeric** $f_i$, $F$, $t_{dispatch}$, $p$, $n$
**Boolean** sat
**Input** $f_1 .. f_{10}$, $t_{dispatch}$
**Output** sat
**Initialization PreCondition**
$\neg Fulfill(Determine\ t_{dispatch}\ to\ make\ p > 50\%\ and\ n < 350)$
**Initialization TriggerCondition**
$I_sensor_i.value \neq ""$
**Initialization PostCondition**
$input = output(I_sensor_i.value)$
**Fulfillment PreCondition**
$sat = ""$
**Fulfillment TriggerCondition**
$Verification(p(t_{dispatch}, F) > 50\% \wedge n(t_{dispatch}, F) < 350)$
**Fulfillment PostCondition**
$sat = Return\ value\ of\ Verification \wedge output(sat)$

Fig. 13. Specification of *A1*

**Plan** Decide $t_{dispatch}$ to hold *p*>50% and *n*<350
**From Adaptive Goal** Determine $t_{dispatch}$ to make *p*>50% and *n*<350
**Attribute**
**Numeric** $F$, $t_{dispatch}$, $t^{new}_{dispatch}$
**Boolean** sat
**Input** sat, $F$, $t_{dispatch}$
**Output** $t^{new}_{dispatch}$
**Initialization PreCondition**
$\neg Fulfill(Determine\ t_{dispatch}\ to\ make\ p > 50\%\ and\ n < 350)$
**Initialization TriggerCondition**
$input = 0$
**Initialization PostCondition**
$t^{new}_{dispatch} = ""$
**Fulfillment PreCondition**
$sat = ""$
**Fulfillment TriggerCondition**
$while(sat \neq 1)$
$\{t^{new}_{dispatch} = t_{dispatch} + 1;$
$Verification(p(t^{new}_{dispatch}, F) > 50\% \wedge n(t^{new}_{dispatch}, F) < 350)$
$t_{dispatch} = t^{new}_{dispatch}\}$
$return\ t_{dispatch}$
**Fulfillment PostCondition**
$t^{new}_{dispatch} = t_{dispatch} \wedge output(t^{new}_{dispatch})$

Fig. 14. Specification of *P1*

## V. Requirements-driven Adaptation Mechanism

This section presents how to get adaptation mechanisms from AGM specifications. Adaptation mechanisms are composed of four parts: monitoring, analyzing, planning and executing, which are also known as adaptive tasks. We give the generic algorithm based on the classification of violations in Section IV-B.

Monitoring algorithm should provide the ability of monitoring context uncertainty and components uncertainty. The variables and sensors are specified in specifications (Figure 12).

**Algorithm 1** Monitor Context and Components Uncertainty

```
monitorUncertainty (AGM specifications){
  for each Monitor task M{
     select M.Specification.Attribute.Numeric.variables
                                          as monitored variables
     select M.Specification.Attribute.Class.instance as monitor
     gauge values V
     return V
}}
```

Analyzing algorithm is used for diagnosing whether the FR or NFR are violated. The inputs of analyzing algorithm are specifications of *Analyze* task (figure 13) and the returned value of variables from monitoring algorithm.

**Algorithm 2** Diagnose Requirements Violations

```
diagnoseViolations (AGM specifications, monitored value){
  for each Analyze task A{
  violationType VT
  if A.InitializationPostCondition=TRUE
     Match violation type
     if (father Ag is affected by context uncertainty and Ag is
          related to FR)
        { computing and verifying Ag.Invariant
          if (Ag.Invariant=FALSE)  VT=ConU_FR
          else  VT=None
          return VT}
    end if
     if (father Ag is affected by context uncertainty and Ag is to
           with sg)
        {  computing sg.Attribute.Numeric.Utility
           if(Utility<desired value)  VT=ConU_NFR
           else  VT=None
           return VT}
    end if
     if (father Ag is affected by components uncertainty and Ag
          is to related with FR)
        {  if (monitored value=" ")  VT=ComU_FR
           else  VT=None
           return VT}
    end if
     if (father Ag is affected by components uncertainty and Ag
          is related to NFR)
        { if (detect noise=TURE)  VT=ComU_NFR
           else  VT=None
           return VT }
    end if
  end if
}}
```

Planning algorithm is designed for determining how to reconfigure the system for mitigating these violations. The inputs are specifications of *Plan* task (Figure 14) and the violation type.

**Algorithm 3** Determine reconfiguration

```
decisionMaking (AGM specifications, violation_type){
  for each Plan task P{
    if P.InitializationPostCondition=TRUE
       Match violation type
       if (violation_type = None)  return currentconfiguration
       if (violation_type = ConU_FR or ConU_NFR)
         {while (violation_type≠None)
            do {adjust P.Attribte.Numeric.parameterValue
                diagnoseViolations }
            end do
          return new parameterValue }
       end if
       if (violation_type = ComU_FR or ComU_NFR)
         {while (violation_type≠None)
            do { replace failed components with
                      M.Specification.Attribute.Class.instance
                 diagnoseViolations }
            end do
          return new structureconfiguration }
      end if
    end if
}}
```

Executing algorithm is built for carrying out the reconfigurations derived in planning algorithm. The reconfiguration can either be parametric or structural.

**Algorithm 4** Execute Reconfiguration

```
executeDecision (AGM specifications, reconfiguration){
  for each Execute task E{
    if  E.InitializationPostCondition=TRUE
      if (reconfiguration= new  parameterValue)
         P.Attribte.Numeric.parameterValue=
                                   new parameterValue
      end if
      if ( reconfiguration= new structureconfiguration)
        M.Specification.Attribute.Class.instance=
                                   new structureconfiguration
      end if
    end if
}}
```

Thus, we derive the generic adaptation mechanism algorithms according to AGM specifications. In next Section, we conduct two simulation experiments to demonstrate the adaptation mechanism algorithms' ability.

## VI. Experimental Evaluation

To evaluate the effectiveness of our approach, we conduct two experiments based on the scenarios of HRCS on a computer with Intel Core 3110M CPU and 2G memory. HRCS is modeled by using AnyLogic. Road Traffic Library and Railway Library are used for runtime simulation. The first experiment focuses on mitigating violation of R2. We present how to trade-off between softgoal *sg2* and *sg3* by parametric adaptation. In the second experiment, we illustrate how to reconfigure the system for mitigating violation of functional requirements R3 and R4.

### *A. Mitigating Violation of Non-functional Requirements*

For diagnosing violation of R2, we suggest leveraging utility-based quantitative verification. Utility can be used to represent the satisfaction degree of NFR, just as what we state in Section IV-A. First, we give the function relation of these utility and variables.

$$U_E = \begin{cases} 1 & E > 20 \\ 0 & E \le 20 \end{cases} \tag{1}$$

$$t_{open} \begin{cases} = 4 & U_E > 0 \\ \in [4, 7) & U_E \le 0 \end{cases} \qquad \mathrm{U_{open}} = \frac{7 - t_{open}}{3} \tag{2}$$

$$t_{close} \begin{cases} = 4 & U_E > 0 \\ \in (1, 4] & U_E \le 0 \end{cases} \qquad \mathrm{U_{close}} = \frac{t_{close} - 1}{3} \tag{3}$$

$$\mathrm{U_{safety}} = \mathrm{U_E} + \frac{1}{2}(1\text{-}\mathrm{sgnU_E})\left|\mathrm{U_{open}} + \mathrm{U_{close}}\text{-}2\right| \tag{4}$$

$$\mathrm{U_{pass}} = \frac{\mathrm{U_{open}} + \mathrm{U_{close}}}{2} \tag{5}$$

Equation 1 depicts utility function of illuminance is 0-1 function. We assume the optimal close/open time interval are both 4s when E>20. When E<20, the time interval can be adjusted according to Eq. 2 and Eq. 3. In Eq. 4, $\mathrm{sgnU_E}$ is the symbolic function of $U_E$. When E>20, $U_{safety}$ equals to $U_E$. That is to say under better illumination, $U_{safety}$ is not affected by $t_{close}$ and $t_{open}$. When $U_E$ decreases to 0, $U_{safety}$ can be tuned by adjusting $t_{close}$ and $t_{open}$. $U_{pass}$ means that the more time vehicles can have to pass the crossing, the larger value it will be.

Integrate Eq. 4 and Eq. 5, we can derive

$$U_{wait} + U_{safety} = 1 \tag{6}$$

$$MRS_{U_{safety}U_{wait}} = -\frac{\Delta U_{pass}}{\Delta U_{safety}} = 1 \tag{7}$$

MRS refers to marginal rate of substitution. Equation 7 implies that time resource distributed into safety efficiency and pass efficiency result in *Pareto Optimality*. That means our assumption is reasonable. Under *Pareto Optimality*, we can better trade-off between the two "commodities". When *E*<20, we assume the desired $U_{safety}$ is above 0.7. The dynamic adaptation process is shown in Figure 15.

X-axis refers to time. Illuminance is monitored under 20lx and $U_{safety}$ is diagnosed to be 0 at 9, 13 and 18 time point. System first tries to tune $t_{close}$ and $t_{open}$ once violation occurs. However, with iterative verification, R4 is still dissatisfied. Then system tune $t_{close}$ and $t_{open}$ for the second time at 10, 14 and 19 time point. We can see $U_{safety}$ of the former two cases are above 0.7, which means R2 hold at 10 and 14 time point. Thus the adaptation is achieved. For the third case, further tuning is needed.

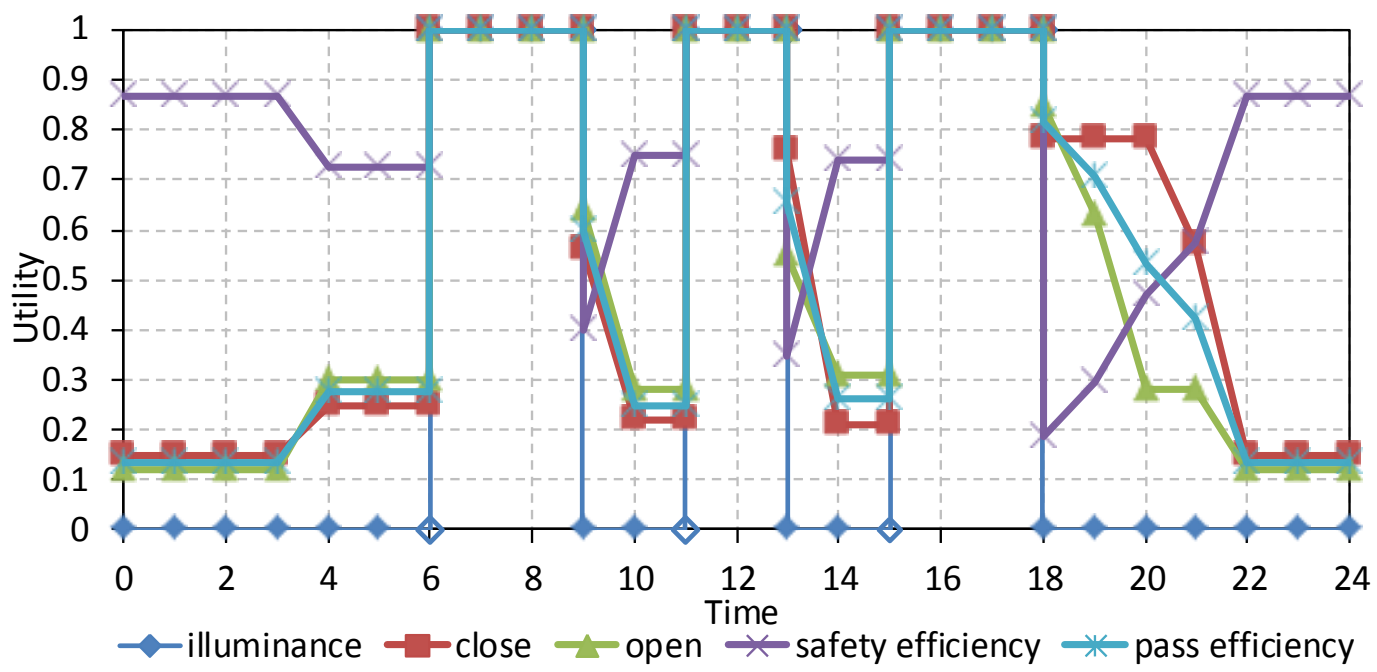

Fig. 15. Adaptation of NFR violation

## B. Mitigating Violation of Functional Requirements

This experiment mainly presents the adaptation of R4 and R5. We design two comparison experiments. The context uncertainty is simulated through changes of vehicle flow.

TABLE IV. CONFIGURATION OF PARAMETERS

| Comparison Experiment | Configuration of Parameters | | | |
|---|---|---|---|---|
| | ***Vehicle Flow from North*** | ***Vehicle Flow from South*** | $t_{close}$ | $t_{open}$ |
| 1 | 15Vehicles/m | 18 Vehicles/m | 4s | 4s |
| 2 | 18Vehicle/m | 20 Vehicles/m | 4s | 4s |

In experiment 1, $t_{dispatch}$ is set to 5min. The x-axis of Figure 16(a) refers to driving time, while y-axis refers to the percentage of vehicles. The x-axis of Figure 16(b) refers to virtual time, while y-axis refers to the number of vehicles on highway. In Figure 16(a), the yellow histogram refers to the vehicles from south to north, while the blue histogram refers to those from north to south. The green vertical line refers to the mean of driving time from south to north while the red vertical line refers to the mean of driving time from the other direction. Based on the statistical result of 165118 samples, we can calculate $p_{blue}$=66.84% and $p_{yellow}$=80.04%, so R3 hold. Besides, the cureve in Figure 16(b) depicts R4 holds.

In experiment 2, the vehicle flow at both sides of highway is monitored increasing by Infrared sensors, while $t_{dispatch}$ is also 5min. We can derive the result in the same way. In Figure 17, $p_{blue}$=38.77%, $p_{yellow}$=48.51% and n>350. Thus, both R3 and R4 are diagnosed violated.

According to the adaptation mechanism algorithm, we should tune the parameter $t_{dispatch}$ at runtime by increasing its value. Do $t_{dispatch}^{new} = t_{dispatch} + 1 = 6s$ and verify R3 and R4 again. The result is presented in Figure 18 with $p_{blue}$=90.57%, $p_{yellow}$=87.45% and n<350. Thus, R3 and R4 hold again and adaptation is accomplished.

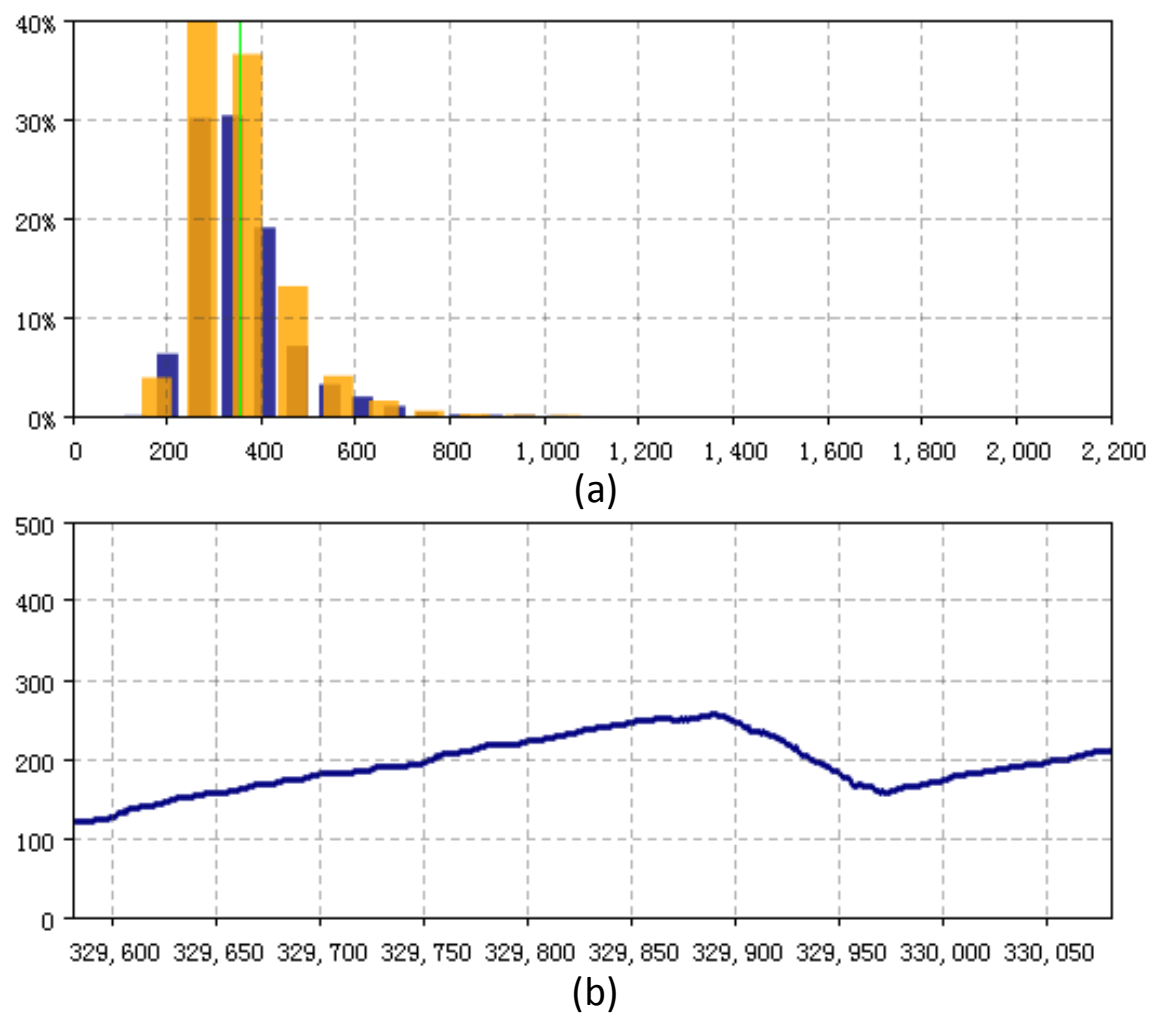

Fig. 16. Histogram of driving time (a) and vehicle amount curve (b) in experiment 1 when tdispatch=5s and R3, R4 hold

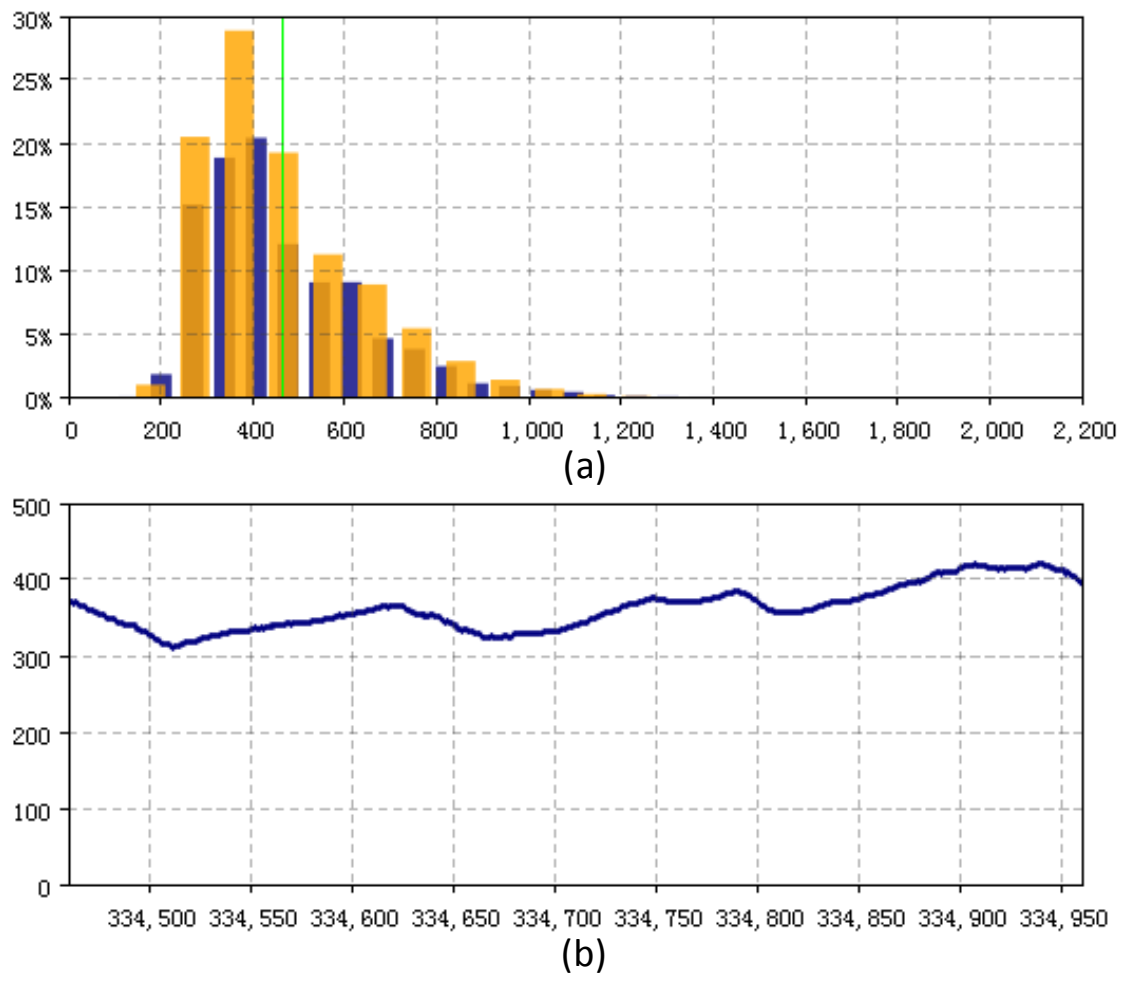

Fig. 17. Histogram of driving time (a) and vehicle amount curve (b) in experiment 2 when tdispatch=5s and R3, R4 are violated

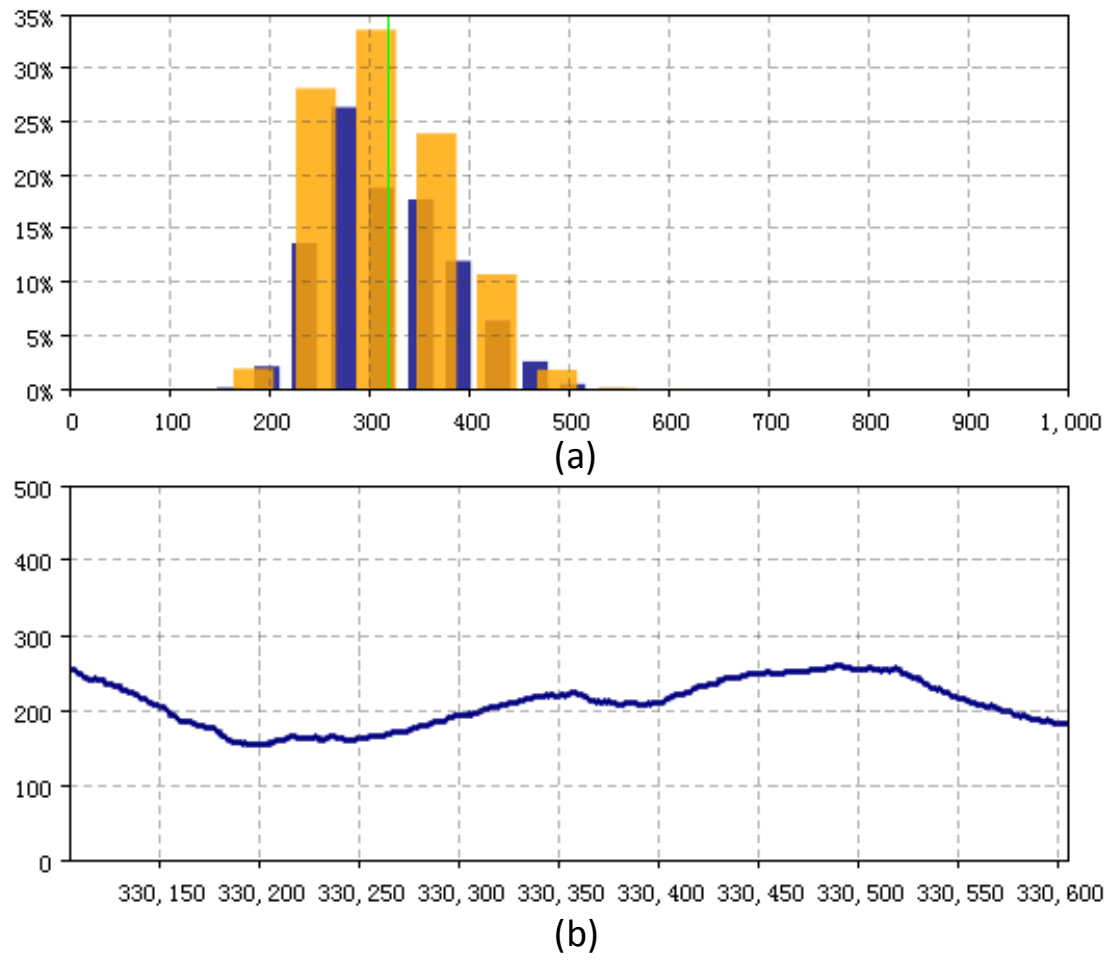

Fig. 18. The result after parametric adaptation

## VII. Related Work

Over the past decade, researches and practitioners have developed a variety of methodologies, frameworks, and technologies intended to support building adaptation.

**Requirements model.** Wittle et al. [4] introduced RELAX, a formal requirements specification language that leverage Fuzzy Branching Temporal Logic to specify the uncertain requirements of self-adaptive systems. Our specification work differs from theirs in that they just specify the static properties of requirement, e.g. involved environment and relations with other entities. However, AGM specification can describe the dynamic properties of requirements, e.g. the initialization and fulfillment of goals or tasks, input and output of tasks, the variant of goals. In a subsequent work [5], Cheng et al. extended RELAX with goal modeling to specify uncertainty in the goals, while our AGM can not only describes uncertainty, but also describes how to mitigate uncertainty by refining adaptive goal with MAPE loop. There are some other works on addressing uncertainty. Goldsby and Cheng [6] presented behavior models to deal with uncertainty in environment through processes in model-driven engineering. Our REDAPT is also a model-driven method, but we focus both on environment and system itself. To managing requirements time uncertainty, Salay et al. [7] proposed partial models to represent uncertainty in requirement and illustrate uncertainty reduction by reason through the traceability relations.

**Monitoring and Diagnosing.** Requirement monitoring aims to track systems' runtime behavior for detecting requirements violations. Diagnosing always comes along with monitoring. Fickas and Feather proposed goal-oriented methods for runtime requirements monitoring [8] and mechanisms for repairing deviations caused by unsatisfied domain assumptions [9]. However, our work also focuses on how to monitor the deviation of system itself. Besides, REDAPT gives an answer to the question about how to provide monitoring specifications [8] [10]. Another representative work on monitoring is requirements awareness. Requirements-aware approaches [11-14] considered requirements as runtime entities and monitoring as meta-requirements about whether the requirements hold during runtime. Our work not only treats requirements as runtime entities, but also treat requirements model as runtime entities, i.e. requirements models at runtime. Besides, Ramirez and Cheng [15] proposed to use utility changes to depict requirements violation, while we just leverage utility for represent satisfaction degree of NFR.

**Achieving adaptation.** Notable work of Baresi et al. [16] introduced adaptive goal into KAOS model and provided adaptation mechanism by adopting fuzzy membership function. Our work differs from theirs in that adaptive goals in FLAGS refers to the goals that can be fuzzy or modified, while adaptive goals in REDAPT represent the adaptation needs of both NR and NFR. In addition, adaptive goals in FLAGS couldn't adjust themselves online, because it has no online adaptation mechanisms. On the contrary, REDAPT provides plenty and generic adaptation mechanisms for supporting adjustment at runtime. Wang et al. [17] [18] presented goal-oriented methods for diagnosing and repairing requirements violations by selecting the optimal structural configuration that contributes most positively to system's NFR. Our work not only focuses on structural adaptation but also parametric adaptation. Esfahani et al. [19] provided POISED for tackling the challenge posed by internal uncertainty by introducing possibility theory to quantify uncertainty. While REDAPT introduces utility functions to quantify satisfaction of NFR.

**Architecture-based methods.** Researches also endeavor to develop architectural methods and techniques. Cheng and Garlan [20] described three specific sources of uncertainty and address the way of mitigating them in Rainbow framework [21]. In notable work of Oreizy et al. [22], A broad framework for studying and describing evolution is introduced by addressing several issues on architectural, that serves to unify the wide range of work in the field of dynamic software adaptation. Vogel and Giese [23] proposed a model-driven approach to provide multiple architectural runtime models at different levels of abstraction as a basis for adaptation.

## VIII. Conclusion

This paper presents REDAPT, short for Requirements-Driven adaptation, a new method of mitigating runtime uncertainty for self-adaptive systems. The contributions of REDAPT are as follows. First, by developing adaptive goal model (AGM), the approach provides generically comprehensive processes for reflecting Problem Layer into Model Layer through modeling adaptive goals, context uncertainty, components uncertainty and refining adaptive goals with MAPE loops. Second, by specifying elements of AGM, REDAPT presents the dynamic properties of AGM that bridge the gap between Model Layer and Adaptation Layer. Third, by proposing adaptation algorithm, we can achieve dynamic adaptation for mitigating runtime uncertainty through monitoring, analyzing, planning and executing. Thus, the adaptation problem returns to Problem Layer again. The three layers compose a feedback loop that controls the adaptation of self-adaptive systems.

In our future work, we will concentrate on investigating appropriate behavior models at runtime for representing self-adaptive systems and transitions from our AGM to behavior model. In parallel, we will also intend to provide novel adaptation mechanisms, especially runtime verification and decision making.